\documentclass[11pt, oneside]{amsart} 
\usepackage{geometry}
\usepackage{amsmath,amssymb,mathtools,amsthm,bbm}

\usepackage{times}
\usepackage{bm}
\usepackage{xcolor}
\usepackage{adjustbox,caption,subcaption,booktabs,multirow,bigints}

\def\T{{ \mathrm{\scriptscriptstyle T} }}

\def \tr{\mathrm{trace}}

\newtheorem{theorem}{Theorem}[section]
\newtheorem{lemma}[theorem]{Lemma}

\begin{document}




\title{Divide-and-Conquer for Covariance Matrix Estimation}

\author{ {\bf Gautam Sabnis},\\
Department of Statistics, Florida State University, Tallahassee, FL, \\email: gsabnis@stat.fsu.edu
\\{\bf Debdeep Pati}, \\Department of Statistics, Florida State University, Tallahassee, FL, \\email: debdeep@stat.fsu.edu
\\{\bf Barbara Engelhardt}, \\Computer Science Department, Princeton University, Princeton, NJ, \\email: bee@princeton.edu
\\{\bf Natesh S. Pillai} \\Department of Statistics, Harvard University, Cambridge, MA, \\email: pillai@fas.harvard.edu}

\maketitle

\begin{abstract}

We propose a distributed computing framework, based on a divide and conquer strategy and hierarchical modeling, to accelerate posterior inference for high-dimensional Bayesian factor models. Our approach distributes the task of high-dimensional covariance matrix estimation to multiple cores, solves each subproblem separately via a latent factor model, and then combines these estimates to produce a global estimate of the covariance matrix. Existing divide and conquer methods focus exclusively on dividing the total number of observations $n$ into subsamples while keeping the dimension $p$ fixed. Our approach is novel in this regard: it includes all of the $n$ samples in each subproblem and, instead, splits the dimension $p$ into smaller subsets for each subproblem. The subproblems themselves can be challenging to solve when $p$ is large due to the dependencies across dimensions. To circumvent this issue, we specify a novel hierarchical structure on the latent factors that allows for flexible dependencies across dimensions, while still maintaining computational efficiency. Our approach is readily parallelizable and is shown to have computational efficiency of several orders of magnitude in comparison to fitting a full factor model. We report the performance of our method in synthetic examples and a genomics application.
\end{abstract}
\normalsize
\section{Introduction} \label{intro}
Factor models attempt to characterize the covariance structure among a large number of random variables by identifying common sources of variation and separating these from idiosyncratic, variable-specific noise. The common sources of variation are assumed to be captured in a small number of unobservable factors. These models have numerous applications spanning a broad range of fields, including portfolio allocation and risk management \cite{fan2013large}, high-dimensional classification \cite{friedman2001elements,shao2011sparse}, studying climate interactions \cite{bickel2008covariance}, controlling false discovery rates in multiple hypothesis testing \cite{efron2010correlated,fan2012estimating}, and gene expression studies \cite{West03bayesianfactor,lucas2006sparse,carvalho2012high,bhattacharya2011sparse, touloumis2015nonparametric,zhao2016bayesian}. 

Motivated by applications in gene expression studies, we focus on Bayesian latent factor models \cite{West03bayesianfactor}, where the dependencies among the high-dimensional observations are explained through a smaller number of common, sparse, latent factors. The methodology outlined in \cite{West03bayesianfactor} crucially exploits \emph{sparsity} and has been successfully employed in many scientific applications \cite{mangravite2013statin,West03bayesianfactor,lucas2006sparse,carvalho2012high,pati2014posterior}. Bayesian methods also have the advantage of automatic tuning of hyperparameters and uncertainty quantification through the posterior distribution. In the last decade, several shrinkage priors \cite{West03bayesianfactor,lucas2006sparse,bhattacharya2011sparse,carvalho2010horseshoe,pati2014posterior} have been proposed to induce sparsity on the factor loadings. 
In the ``large p, small n'' scenario, the shrinkage priors developed in \cite{pati2014posterior} are also proved to have attractive theoretical properties.
A key computational challenge encountered while implementing these methods, particularly when dealing with massive covariance matrices, is that they require storage of large matrices in memory and repeated, computationally expensive matrix inversions. These issues severely limit the scalability of most of the above methods even in moderate dimensions. The main goal of this paper is to have a modeling framework to remove the computational bottlenecks that arise in posterior inference for high dimensional 
sparse latent factor models.\par
To improve computational tractability and leverage the growing availability of platforms for distributed computing, we propose a divide and conquer approach for high-dimensional covariance matrix estimation. At a high level, our approach randomly divides the high-dimensional data into low dimensional subproblems, solves these subproblems in parallel using existing Markov chain Monte Carlo methods, and combines these estimates via a two-level hierarchical model to produce a global estimate for the covariance matrix. 

The idea behind the this framework is pervasive in the computer science, referred to as \emph{parallel and distributed computing} \cite{andrews1999foundations}. \cite{mackey2011divide} proposes a divide-factor-conquer framework for recovering a matrix factorization that randomly divides the original matrix factorization problem into smaller submatrices, solves the problem for each submatrix, and then combines the solutions to each submatrix problem using efficient techniques from randomized matrix approximation. 

We call attention to some recent related works using a divide and conquer type strategy. In \cite{zhang2013divide}, the authors establish optimal convergence rates for a decomposition-based scalable approach to kernel ridge regression. In \cite{minsker2014robust}, the authors use the Weierstrass transform to combine subset posterior estimates by running independent MCMC chains for each data subset. Recently, \cite{cheng2015computational} explored the statistical versus computational trade-offs to find the computational limits of divide and conquer method in a regression setup. Our approach relies on the same basic divide and conquer ideas as the related methods, but fundamental differences with our approach distinguish it from previous work. Notably, in \cite{zhang2013divide,minsker2014robust,cheng2015computational}, the authors focus on the ``large $n$" problem, where the the samples are assumed to be independent and identically distributed, $n \gg p$, and subsetting occurs in the number of samples $n$ and not in $p$. Our approach is starkly different from previous work in that we split dimension $p$ across the subproblems.  The main modeling challenge in splitting $p$ features across subproblems is to have a flexible framework to capture dependencies across dimensions, during the ``conquer step".  The very recent work \cite{li2015embracing} proposes a similar divide and conquer scheme that splits across the dimensions and utilizes the variables in each group to estimate latent factors. The estimator proposed in  \cite{li2015embracing} is a simple average across of estimators from different subproblems. From the Bayesian point of view, this might not adequately capture posterior uncertainty due to the dependencies across different dimensions. 



\section{Latent Factor Models In High Dimensions} \label{sec:factor}
Let $y_i$ be a $p$-dimensional zero-mean, normal distributed random vector with covariance matrix $\Sigma$. A $k$-dimensional $(k < p)$ latent factor model for
$y_i$ may be expressed as
\begin{equation}  \label{factor}
y_i = \Lambda{\eta_i} + \epsilon_i,  \quad i = 1, \dots, n
\end{equation} 
where $\Lambda \in {\mathbb{R}}^{p \times k}$ is a matrix of unknown factor loadings, $\eta_i \in {\mathbb{R}}^{k}$ is a vector of latent factor scores with $\eta_i \sim \mathrm{N}(0,\mathrm{I}_p)$, and $\epsilon_i$ is a $p$-dimensional vector of independent, idiosyncratic noise: $\epsilon_i \sim N_p(0, \Omega)$ and $\Omega = {\text{diag}}(\sigma_1^{2}, \dots, \sigma_p^{2})$. The factor scores $\eta_i$ are assumed to be independent of noise terms, and thus the covariance structure of $y_i$ admits a factor decomposition of the form,
\begin{equation} \label{decomp}
\Sigma = \Lambda{\Lambda}^{T} + \Omega
\end{equation} 
explicitly separating the commonalities ($\Lambda\Lambda^{T}$) from specificities ($\Omega$) in the variation of $y$. This reduces the number of parameters to be estimated from $O(p^2)$ parameters in an unstructured covariance matrix to $O(pk + p)$ parameters in \eqref{decomp}. Factor models thus provide a convenient and parsimonious framework for modeling covariance matrices, particularly in applications with moderate to large $p$.   
The factor model in \eqref{factor}, without further constraints, is non-identifiable. Assuming $\Lambda$ to be lower triangular eliminates the identifiability issues in \eqref{decomp} \cite{geweke1996measuring}. However, one does not require the identifiability of the loading elements for a wide class of applications, including covariance estimation, variable selection, and prediction. A standard Bayesian approach involves placing a prior distribution on $(\Lambda, \Omega)$ and learning the number of factors $k$ on the fly \cite{bhattacharya2011sparse,knowles2011nonparametric}. 

Although the original specification of the factor model reduces the number of parameters from quadratic to linear in $p$, the estimation problem is still challenging when $p \gg n$. To address this issue, \cite{West03bayesianfactor} introduced sparse latent factor models to allow many of the loadings to be exactly zero by placing a point mass mixture prior having a probability mass at zero. Although sparsity favoring priors have been successfully implemented in genomic applications \cite{carvalho2012high,lucas2006sparse} and shown to enjoy appealing theoretical properties \cite{pati2014posterior}, posterior computation under such priors can be daunting in high-dimensional cases. This is mainly due to complexity associated with 
repeated inversion of $k \times k$ matrices, which can be intractable when $p$ is large. Moreover, there are substantial costs involved with storing $p \times k$ dimensional factor loading matrices in memory as the Gibbs sampler proceeds.
\section{The divide and conquer Framework} \label{dfc}
In this section, we present our distributed computing strategy for covariance matrix estimation in the Bayesian latent factor model setting \eqref{factor} with a generic shrinkage prior on the factor loadings matrix. Assume that we have $g \geq 1$ cores at our disposal, the algorithm proceeds as follows:  \\
{\em D-Step (Divide across dimension):} Randomly partition $y_i$ into $g$ $p_{g}$-dimensional sub-vectors, $\{{{y}}^{(1)}, \dots, {{y}}^{(g)}\}$ where ${{y}}_{i}^{(m)} \in {\mathbb{R}}^{p_g}, \;\; {m} = 1, \dots, g \; \mbox{and} \; p_g = p/g$.  
For simplicity, we assume that $p$ is a multiple of $g$ and that $p_g = p/g$. For arbitrary $p$ and $g$, $p$ can always be partitioned into $p_g$ subvectors, each with either $\lfloor {p/g} \rfloor$ or $\lceil {p/g} \rceil$ elements. \\
{\em F-step (Obtain individual fits):} We model the sub-vectors $y_i^{(m)}$, using (\ref{factor}), for each $m = 1, \ldots, g$ 
\begin{eqnarray*}
y_i^{(m)} = \Lambda^{(m)} \eta_i^{(m)}  + \epsilon_i^{(m)}, \quad \epsilon_i^{(m)} \sim \mbox{N}(0, \Omega^{(m)})
\end{eqnarray*}
and obtain posterior distribution of $\Sigma^{(m)} \in {\mathbb{R}}^{p_g \times p_g}$ based on a shrinkage prior on $(\Lambda^{(m)}, \Omega^{(m)})$ conditional on the latent factors $\eta_i^{(m)} \in {\mathbb{R}}^{k_g}$. 
This step can be parallelized on $g$ cores.\\
{\em C step (Combine the fits):} This step pools together the posterior samples from $g$ machines and combines the estimates to form a global estimate.  
Since our primary goal is to elucidate the underlying covariance structure, assuming the estimates obtained from different machines are independent ignores the dependence structure of the observed multivariate observations.  In the following, we describe a hierarchical model that induces dependence, through latent factors, among the sub-vectors obtained from the \textit{D-step}.

To facilitate the use of information shared across each sub-estimate, we incorporate a dependency structure between the factors $\eta_i^{(m)}$ across different sub-groups using a data-augmentation technique.  Consider the hierarchical model, 
\begin{equation} \label{hie} 
\eta_i^{(m)} \mid X_i, Z_i^{(m)} = \sqrt{\rho}\;X_i + \sqrt{1 - \rho}Z_i^{(m)}, \;\; i = 1, \dots, n, \;\; m = 1, \dots, g
\end{equation}
where  $X_i \sim \mathrm{N}(0, \mathrm{I}_{k_g})$, is the component of $\eta_i^{(m)}$ that is shared across all the latent sub-factors.  The quantity $Z_i^{(m)} \sim \mathrm{N}(0, \mathrm{I}_{k_g})$ is the component of $\eta_i^{(m)}$ that is idiosyncratic to the specific sub-factor, and $\rho$ is the correlation induced between the latent sub-factors.  A discrete uniform prior on $\rho$ on $[0, 1]$ provides a computational convenient  choice.  Observe that under \eqref{hie}, the marginal distribution of $\eta_i^{(m)}$ is still $\mbox{N}(0, \mathrm{I}_{k_g})$. 
The hierarchical structure described above has two distinct advantages: i) it induces a correlation structure among sub-estimates that is used to combine them in the algorithm, and ii) it does so without increasing the computational complexity of the algorithm.  Using \eqref{hie}, we can rewrite \eqref{factor} for each sub-vector as
\begin{equation} 
y_i^{(m)} = \Lambda^{(m)}\big\{\sqrt{\rho}\;X_i + \sqrt{1 - \rho}\;Z_i^{(m)}\big\} + \epsilon_i^{(m)}, i = 1, \dots, n, \; m = 1, \dots, g.
\end{equation}
The following lemma characterizes the covariance between two sub-vectors $y_i^{(m)}$ and  $y_i^{(m')}$. The proof  is standard and, therefore, omitted. 
\begin{lemma}\label{cov-subgrp}
Let $m, m' \in \{1, \ldots, g\}$ be such that $m \ne m'$. Assume $X_i, Z_i^{(m)} \sim \mathrm{N}(0, \mathrm{I}_{k_g})$ with $Cov\{Z_i^{(m)}, Z_i^{(m')}\} = 0$. Then  $Cov\big\{y_i^{(m)}, y_i^{(m')}\big\}$ = $\rho\Lambda^{(m)}\Lambda^{(m')}$.  
\end{lemma}
The \textit{F-Step} can be efficiently parallelized by running the full conditioning samplers of $\Lambda^{(m)}, \Omega^{(m)}$ on the individual cores, which are then collected to obtain the full conditional distribution of the commonalities $\eta_i^{(m)}$, $\rho$ on a separate core. In our simulations, we observed that the communication cost associated with this step to be insubstantial as compared to fitting a factor model on a single machine.     
In the \textit{C-step}, we form an estimate of the original variance covariance matrix $\Sigma$ using the
following lemma. The global estimate $\Sigma_E$ is obtained by combining the posterior samples from the different cores.    
\begin{lemma} \label{cest}
The estimate for the original covariance matrix $\Sigma$ is obtained using $\Sigma_E = D E D^{\T} + \Omega$, where $D = \text{diag}\big\{\Lambda^{(1)}, \cdots, \Lambda^{(g)}\big\}$, 
$\Omega = \text{diag}\big\{\Omega^{(1)}, \cdots, \Omega^{(g)}\big\}$, $E = I_{k_g} \otimes C$ for a $g \times g$ positive definite matrix $C$ such that  
$C_{mm'} = 1$ if $m=m'$ and $C_{mm'} = {\rho}$ if $m \neq m'$.
\end{lemma}
The proof of Lemma \ref{cest} follows from standard matrix algebra. Observe that  $E \in {\mathbb{R}}^{k \times k}$ consists of $g^{2}$ $k_g\times k_g $- dimensional block matrices.  The final estimator of the covariance matrix obtained in Lemma \ref{cest} is similar in structure to the covariance matrix estimator obtained by fitting a full factor model in (\ref{factor}).
\section{Computational tradeoff of the divide and conquer approach} \label{comp}
Most shrinkage prior distributions on $\Lambda$ are amenable to posterior inference via Gibbs sampling. Each step of the Gibbs sampler requires sampling from the conditional posterior distributions of $\Lambda$ and $\Omega$. These sampling procedures require substantial time and memory constraints for: i) performing several different kinds of matrix operations such as matrix-matrix multiplication, matrix inversion, and Cholesky factorization, and ii) storing large matrices since, for example, the posterior update of $\Lambda$ requires the stored posterior samples of $\eta$, and vice versa. 

While working with the full factor model (\ref{factor}), one iteration of the Gibbs sampler for estimating $\Sigma$ requires $\mbox{O}(k^3 + npk + nk^2 + pk^2)$ floating point operations for matrix computations, and a storage complexity of $\mbox{O}(pk + k^2)$. 
The full conditionals needed for Gibbs sampling using approach are detailed in Appendix 2.
Our method significantly reduces the per-iteration floating point operation complexity to $\mbox{O}({k_g^3 + np_gk_g + n{k_g^2} + {p_g}{k_g^2}})$ operations and storage complexity to $\mbox{O}(p_gk_g + k_g^2)$ on a single machine. Hence, the computational speed up is $g^2$-fold. Such computational gains can also be observed from simulation study results in figures \ref{fig:sub3}, \ref{fig:sub4} and table \ref{tab:3} in \S \ref{simulations}. 

\section{Theoretical properties} \label{theory}
In this section, we investigate to what extent $\Sigma_E  = D ED^{\T} + \Omega $ is a good approximation to $\Sigma = \Lambda \Lambda + \Omega$ where $\Lambda \in \mathbb{R}^{p \times k}$.  The proof of the following Lemmata are deferred to Appendix 1.   
 We first prove that if $\Lambda^{(m)}$ and $\Lambda$ have full column ranks for all $m$ respectively, then the two matrices $D ED^{\T}$ and $\Lambda \Lambda^{T}$ have the same rank. 
\begin{lemma} \label{rank} Suppose $\mbox{rank}(\Lambda^{(m)}) = k_g, m = 1, \ldots, g$ and 
$\mbox{rank}(\Lambda) = k$,  then $A = \Lambda\Lambda^{\T}$ and $A^{*} = D E D^{\T}$ have the same rank. 
\end{lemma}
Under most continuous shrinkage prior distributions on $\Lambda$,  $\mbox{rank}(\Lambda) = k $ with probability one. Hence the approximation $D E D^{\T}$  preserves the rank a posteriori. 
We next show that the prior distribution on $D E D^{\T}$ assigns high probability around $\Lambda_0 \Lambda_0^{\T}$, where each of the columns of $\Lambda_0$ has at most $s < p$ many non zero entries.  We work specifically with the multiplicative gamma process shrinkage prior in \cite{bhattacharya2011sparse} placed on the columns of a generic loadings matrix $\Lambda \in \mathbb{R}^{p \times k}$:
\begin{align} 
\lambda_{jh}\mid{\phi_{jh}},{\tau_{h}} \sim N(0, {\phi_{jh}^{-1}}{\tau_{h}^{-1}}), \quad \phi_{jh} \sim \Gamma(\nu/2, \nu/2), \quad \tau_{h} = {\displaystyle\prod\limits_{l=1}^{h}{\delta_{l}}} \label{eq:MGPS1} 
\\
\delta_1 \sim \Gamma(a_1, 1), \quad \delta_{l} \sim \Gamma(a_2, 1), \quad l \ge 2, \quad {\sigma_{j}^{-2}} \sim \Gamma(a_\sigma, b_\sigma) \quad (j = 1, \dots, p) \label{eq:MGPS2},
\end{align}
where $\delta_l \; (l = 1, \dots, \infty)$ are independent, $\tau_{h}$ is a global shrinkage parameter for the $h^{th}$ column, and the $\phi_{jh}$s are local shrinkage parameters for the elements in the $h^{th}$ column. The $\tau_{h}$s are stochastically increasing under the restriction $a_2 > 1$, which favors more shrinkage as the column index increases.\par
Let $\ell_0[s;p]$ be the space of $s$-sparse vectors $\Lambda_{0h}\in {\mathbb{R}}^{p}$ with $\mid\mbox{supp}(\Lambda_{0h})\mid \le s$ such that $1\le s\le p$ and $s/p \le 1/2$. Also, assume that, for an index set $S$, 
$\Lambda_{hS}$ is the sub-vector of $\Lambda_h$ with elements indexed by $S$. 
The next lemma shows, under the multiplicative gamma process prior, $\mbox{trace}(A^*)$ concentrates around $\mbox{trace}(\Lambda_0 \Lambda_0^{\T})$ with high probability. \par
\begin{lemma} \label{trace}
 Under \eqref{eq:MGPS1}-\eqref{eq:MGPS2}, for any $\epsilon \in (0,1)$,
\begin{multline*} 
\mathrm{pr}(\lvert \tr(A^*) - \tr(\Lambda_0\Lambda_0^{\T}))\rvert < \epsilon) \ge \underset{\tau \in \mathcal{B}}{\inf}\;\mathrm{pr}\bigg\{\left|{\displaystyle\sum\limits_{m=1}^{g}}{\displaystyle\sum\limits_{h=1}^{k_g}}{\lVert {\Lambda_{hS_{0h}}^{(m)}}\rVert}^{2} - {\displaystyle\sum\limits_{h=1}^{k}{\lVert{\Lambda_{0hS_{0h}}}\rVert}^{2}}\right| < \epsilon/2 \mid \tau\bigg\} \\ \exp{}\big[-C\{s\log{}(1/a) + \log{}(2k(p-s)/\epsilon)\}\big] \label{traceconc}
\end{multline*}
for some constant $C > 0$, where 
\begin{align*}
\mathcal{B} = \bigg\{ \tau : \frac{\delta}{da^{s/{p-s}}} < \sqrt{\tau_h} < \frac{\delta}{ca^{s/{p-s}}}\quad \mbox{for every}\, h = 1, \ldots, k\bigg\}
\end{align*}
for $\delta = \sqrt{\epsilon}/\sqrt{2k(p-s)}$.  
\end{lemma}
Lemma \ref{trace} guarantees that, \emph{apriori}, the trace of our estimator concentrates around the trace of the true covariance matrix even for large $p$. This prior concentration is a crucial ingredient for posterior optimality, as outlined in \cite{pati2014posterior}. Although the proof uses the multiplicative gamma process prior specified in \eqref{eq:MGPS1}-\eqref{eq:MGPS2}, a similar concentration result can be obtained for other continuous shrinkage prior distributions in the literature (Refer to Lemma 7.1 of \cite{pati2014posterior}).  When $s$ is significantly smaller than $p$, the second term on the right hand side the inequality, in Lemma \ref{trace}, 
is roughly of the order $\exp\{-s - \log (k p)\}$ and therefore decays slowly. The first term is a small ball probability of smaller dimensional random vectors which is guaranteed to have high concentration. 
Hence, $\mbox{trace}(A^*)$ concentrates around $\mbox{trace}(\Lambda_0 \Lambda_0^{T})$ with high probability. 

 Observe that  $\Sigma_E $  may not be a good approximation to $\Sigma$ element-wise.  However, this does not pose an issue as  in ultra-high dimensions, the full-factor model may not be a good lower dimensional representation of data.  Instead,  the hierarchical model in \eqref{hie} aims to do a localized analysis by decomposing factors as pure (specific to each sub-group) and mixed (shared across  all the sub-groups).  On the other hand, the assurance from Lemmata \ref{rank} and \ref{trace} that the rank and the trace do not significantly deviate from the full factor model helps in calibrating the prior distributions.   
 
 \section{Experimental results on synthetic data} \label{simulations} 
In this section, we explore the performance of our method on simulated data. All simulation experiments were performed in Matlab using a Windows machine with 32GB of memory and a single-threaded 3.2Ghz processor, we simulated data from a factor model $y_i = \Lambda\eta_i + \epsilon_i \; i = 1, \ldots, n$, where $\Lambda_h$ has at most $s$ non-zero elements for $h = 1, \ldots, k$ and $\epsilon_i \sim \mbox{N}(0, \sigma^{2})$ with $\sigma^2 = 0.5$. We randomly allocated the location of the zeros in each column and simulated the nonzero elements independently from $\mbox{Unif}(0.1, 3)$. We divided the task of covariance matrix estimation across $g$ groups on a single machine. 
The local estimators ${\hat{\Sigma}}^{(m)}$ computed for each sub-group were combined to form the global estimator of the covariance matrix as in Lemma \ref{cest}. In our experiments, we evaluated the performance of our method for covariance matrix estimation via latent factor models in terms of computational efficiency and accuracy. For comparison, the operator norm error and computational time associated with fitting a full factor model ($g = 1$) serves as a benchmark in our experiments. 

\begin{figure}[htp!]
\centering
\begin{subfigure}{.4\textwidth}
\centering
\includegraphics[width = 0.9\linewidth]{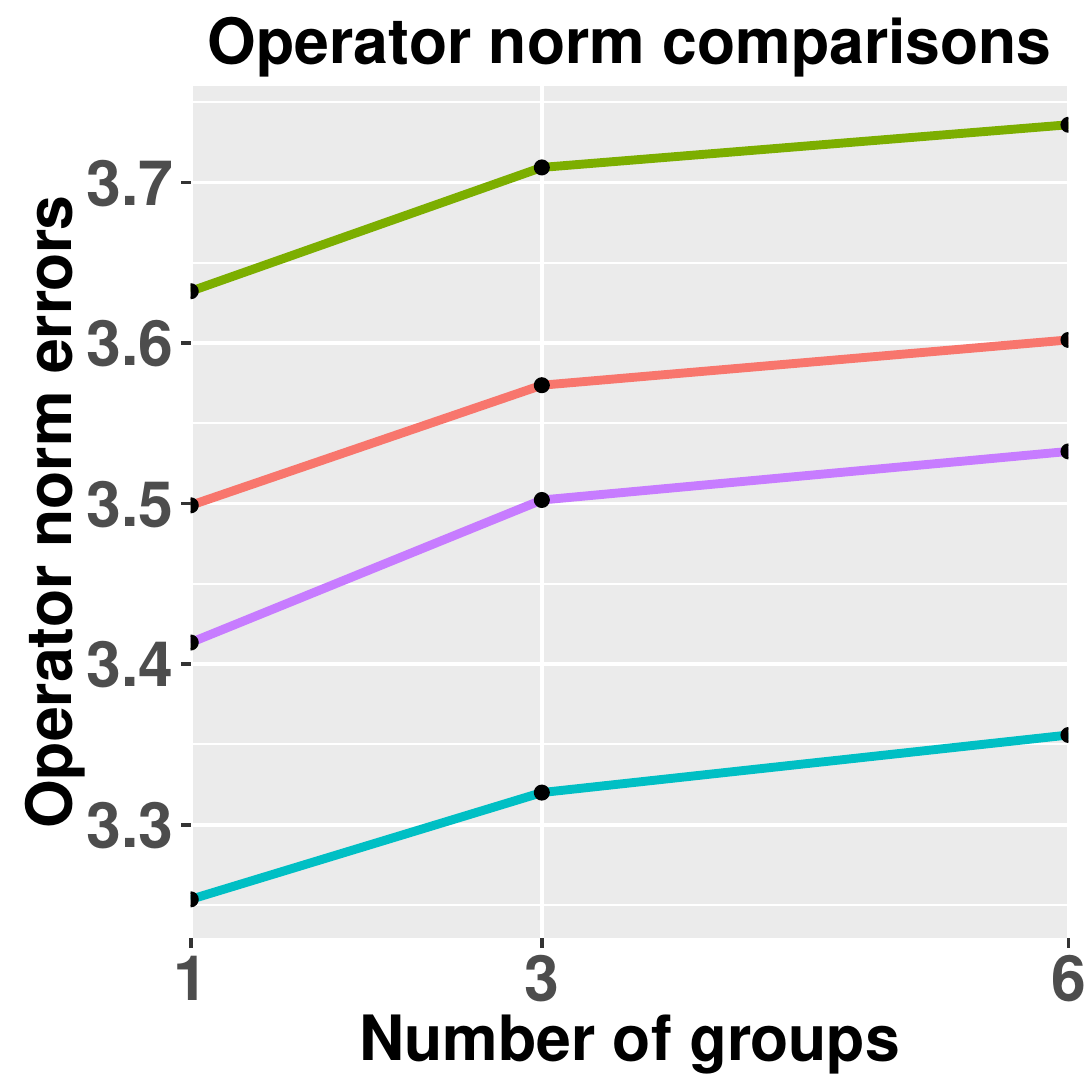} 
\caption{$n$ = 100} 
\label{fig:sub1}
\end{subfigure} 
\begin{subfigure}{.4\textwidth} 
\centering
\includegraphics[width = 0.9\linewidth]{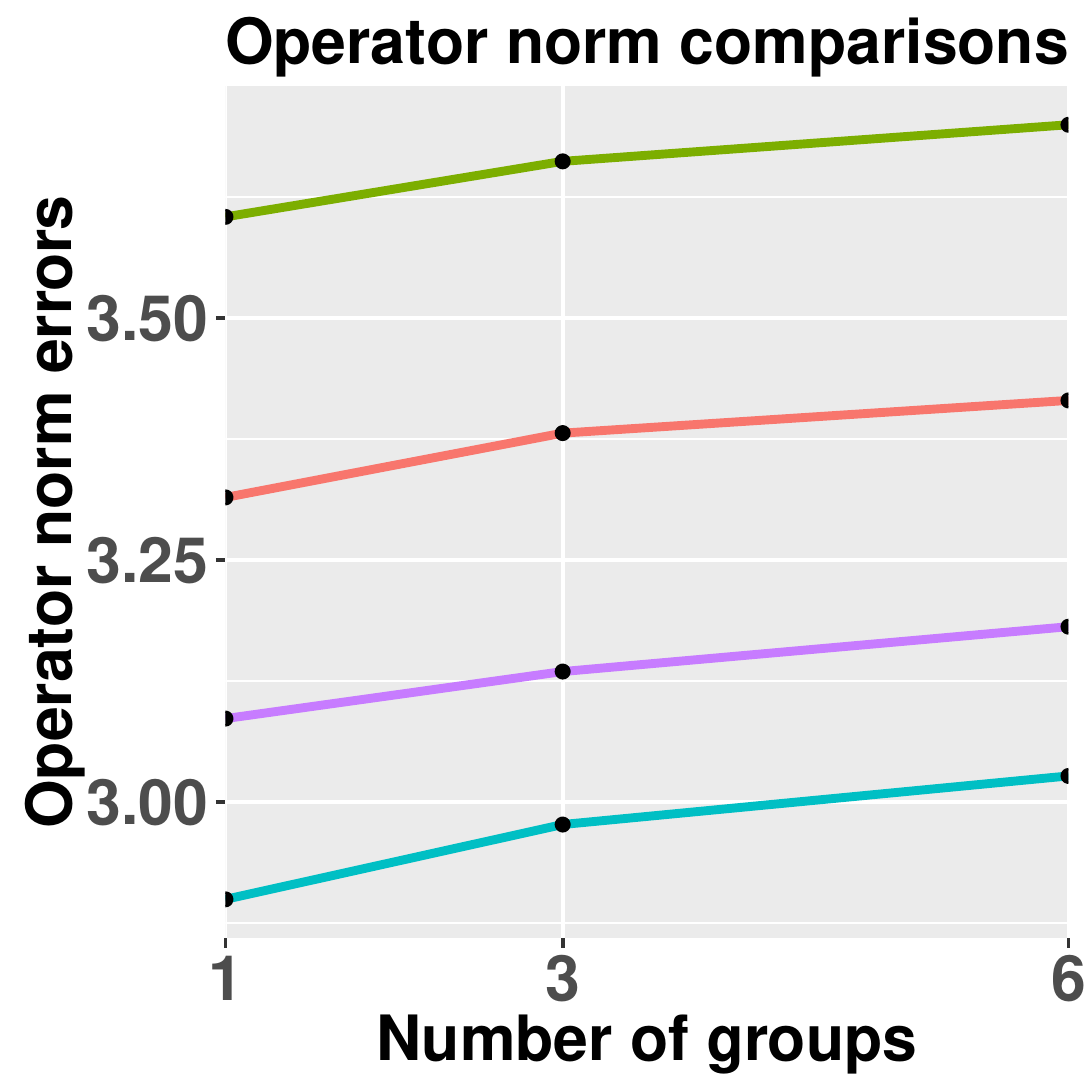} 
\caption{$n$ = 200} 
\label{fig:sub2}
\end{subfigure} 
\caption{A graph comparing the operator norm errors, on a log scale, of the estimator across $g \in \{1,3,6\}$ groups for $p = 252$ (blue), $p = 504$ (purple), $p = 1008$ (red), $p = 2016$ (green)}
\label{fig:test} 
\end{figure}

In Figure \ref{fig:sub1}, we plot the operator norm errors ${\lVert{{\hat{\Sigma}} - \Sigma}\rVert}_{2}$ versus the number of groups $g$, where $g \in \{1, 3, 6\}$. For $p = 2,016$, the operator norm errors on the log scale are given by 3$\cdot$63 for $g = 1$, 3$\cdot$71 for $g = 3$ and 3$\cdot$74 for $g = 6$. In the right panel of Figure \ref{fig:sub2}, we perform an identical experiment, with sample size $n = 200$. For $p = 2,016$ and other values of $p$, we see an improvement in terms of error, (3$\cdot$60, 3$\cdot$68, 3$\cdot$72), for $g \in \{1, 3, 6\}$, respectively, due to increased sample size.  

\begin{figure}[htp!]
\centering
\begin{subfigure}{.4\textwidth}
\centering
\includegraphics[width = 0.9\linewidth]{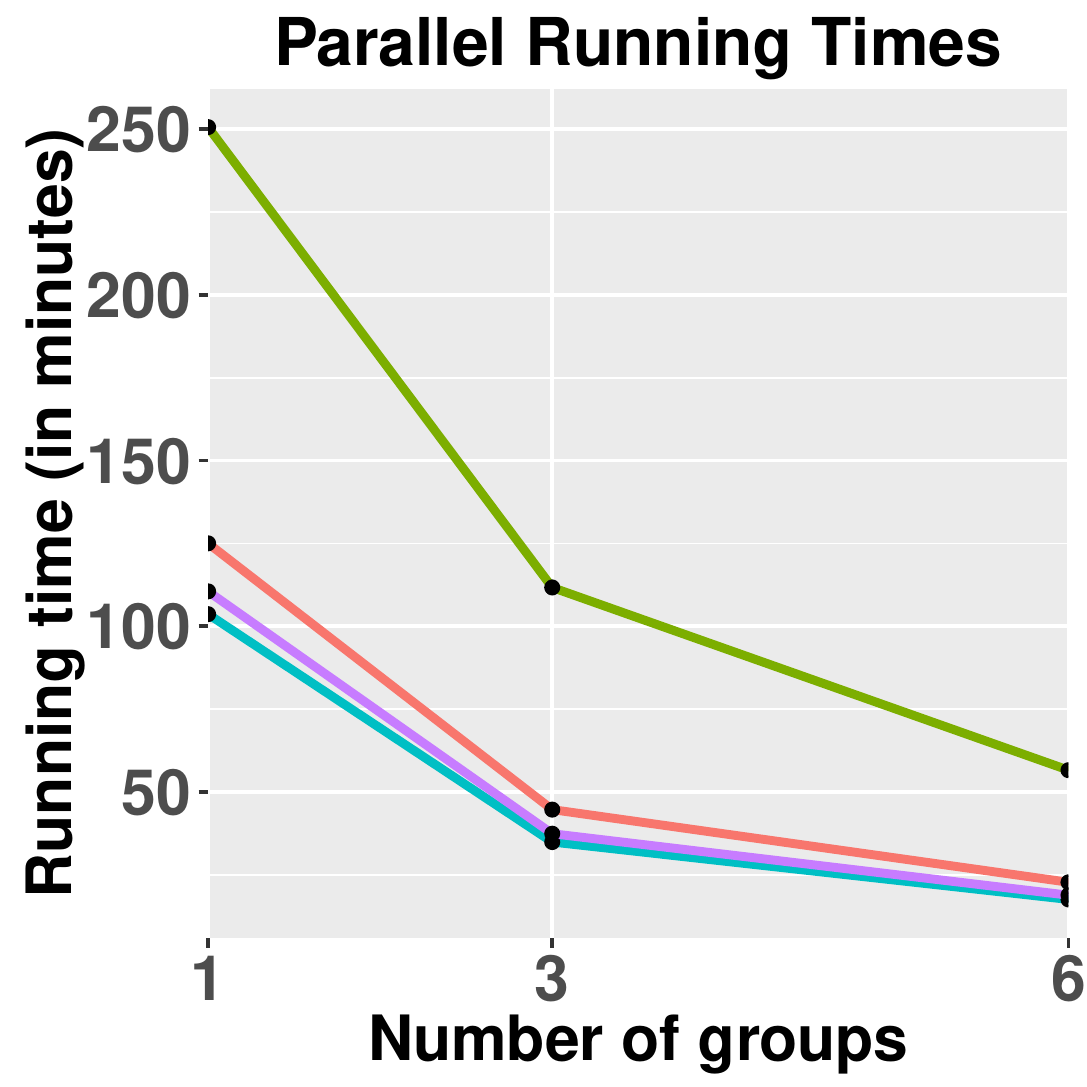} 
\caption{Sample size = 100} 
\label{fig:sub3}
\end{subfigure} 
\begin{subfigure}{.4\textwidth} 
\centering
\includegraphics[width = 0.9\linewidth]{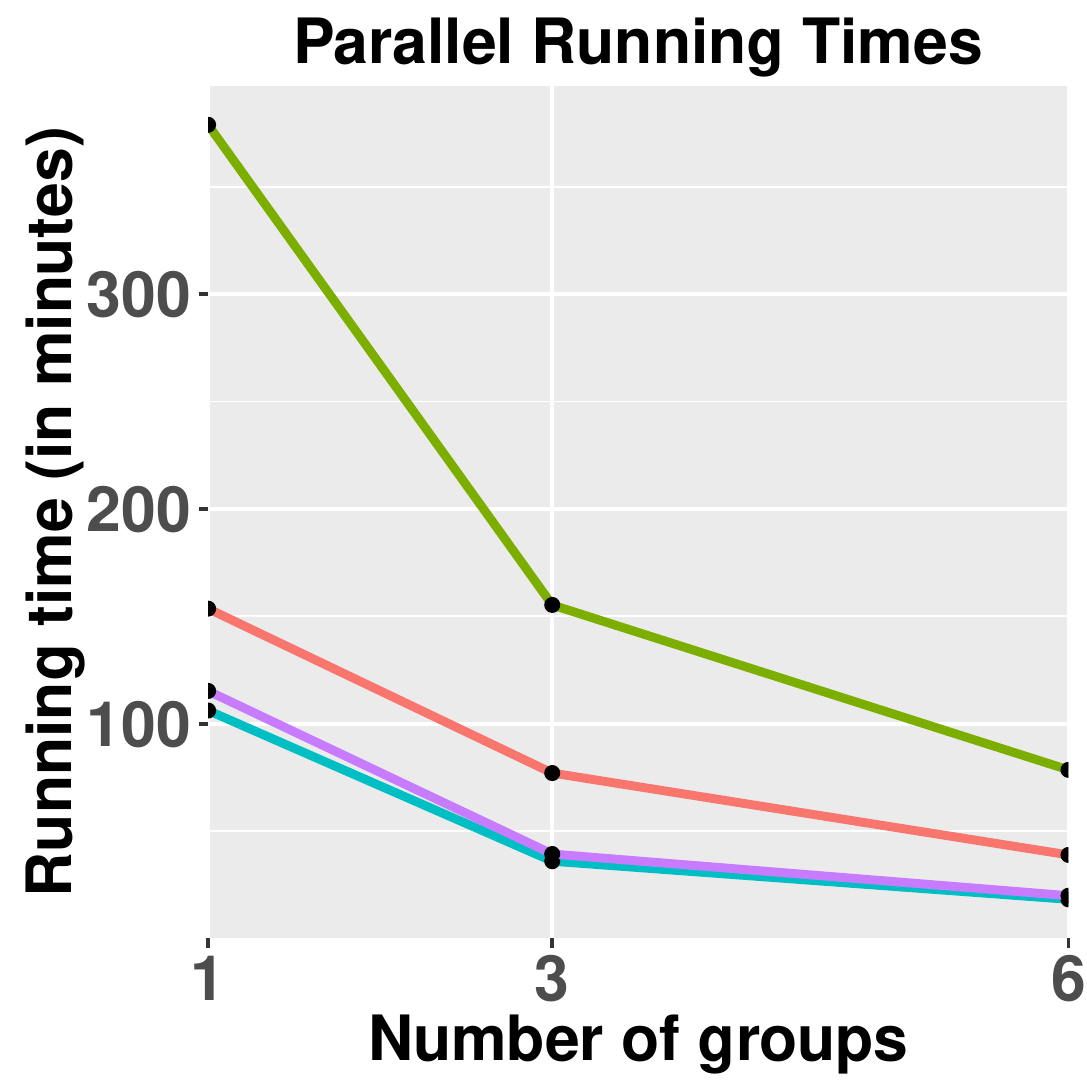} 
\caption{Sample size = 200} 
\label{fig:sub4}
\end{subfigure} 
\caption{A graph showing parallel running time per replicate (in minutes) comparisons for peforming covariance matrix estimation across $g \in \{1,3,6\}$ groups for $p = 252$ (blue), $p = 504$ (purple), $p = 1008$ (red), $p = 2016$ (green)}
\label{fig:test} 
\end{figure}

Figures \ref{fig:sub3} and \ref{fig:sub4} give evidence of the substantial computational tradeoff the strategy offers for the problem of covariance matrix estimation. Here we compare the amount of time (in minutes per replication) required for posterior inference. Note again that $g = 1$ gives the baseline computational time for the task of covariance estimation on a single core.  Again for $p$ as large as $2,016$, we see a $55\%$ reduction in computational time as we divide the task across three machines, and a further reduction of $49\%$ for splitting the task across six machines. This substantial reduction in computational complexity is achieved with minimal cost in error as the figures \ref{fig:sub1} and \ref{fig:sub2} display.  

In Tables \ref{tab:1} and \ref{tab:2}, we report the results of our simulations for four $(p,k)$ combinations with moderate to large $p$, specifically, $(252,6)$, $(504,12)$, $(1008,24)$, and $(2016,36)$. We run the Gibbs sampler for $10,000$ iterations with a burn-in of $4,000$ and collect every $10$th sample to thin the chain. We provide the summaries of mean square error, average absolute bias, and maximum absolute bias for $g \in \{1,3,6\}$ across 20 replicates. The tables show that, without sacrificing accuracy, our method has significant computational benefits. 

\begin{table}[htbp!]
\def~{\hpantom{0}}
\centering
\caption{Comparative performance in covariance matrix estimation in the simulation study. The performance across 20 simulation replicates is reported in terms of running time per replicate, operator norm error , mean squared error ($\times 10^2$), average absolute bias ($\times 10^2$), and maximum absolute bias for different combinations of $p$, $k$, and $g$ with $n = 100$.}
\begin{flushleft} 
\begin{adjustbox}{width = 1\textwidth}
\begin{tabular}{ccccccccccccc} \toprule
{\bf p} & \multicolumn{3}{c}{{\bf{252}}} & \multicolumn{3}{c}{{\bf{504}}} & \multicolumn{3}{c}{{\bf{1008}}} & \multicolumn{3}{c}{{\bf{2016}}} \\ 
{\bf k} & \multicolumn{3}{c}{{\bf{6}}} & \multicolumn{3}{c}{{\bf{12}}} & \multicolumn{3}{c}{{\bf{24}}} & \multicolumn{3}{c}{{\bf{36}}} \\
\cmidrule(lr){2-4} \cmidrule(lr){5-7} \cmidrule(lr){8-10} \cmidrule(lr){11-13} \\
{\bf g} & {\bf 1} & {\bf 3} & {\bf 6} & {\bf 1} & {\bf 3} & {\bf 6} & {\bf 1} & {\bf 3} & {\bf 6} & {\bf 1} & {\bf 3} & {\bf 6}  \\
\hline 
error & 25$\cdot$88 (0$\cdot$34) & 27$\cdot$66 (0$\cdot$06) & 28$\cdot$67 (0$\cdot$00) & 30$\cdot$37 (0$\cdot$30) & 33$\cdot$19 (0$\cdot$02) & 34$\cdot$21 (0$\cdot$00) & 33$\cdot$08 (0$\cdot$32) & 35$\cdot$65 (0$\cdot$07) & 36$\cdot$67 (0$\cdot$00) & 37$\cdot$80  (0$\cdot$65) & 40$\cdot$83 (0$\cdot$2) & 41$\cdot$93 (0$\cdot$00)\\
time & 103$\cdot$65 & 34$\cdot$88 & 17$\cdot$56 & 110$\cdot$53 & 37$\cdot$40 & 18$\cdot$85 & 125$\cdot$03 & 44$\cdot$67 & 22$\cdot$75 & 250$\cdot$62 & 111$\cdot$7 & 56$\cdot$57  \\
mse & 0$\cdot$08 & 0$\cdot$1 & 0$\cdot$5 & 0$\cdot$07 & 0$\cdot$1 & 0$\cdot$3 & 0$\cdot$04 & 0$\cdot$05 & 0$\cdot$1 & 0$\cdot$02 & 0$\cdot$03 & 0$\cdot$07 \\
avgbias & 0$\cdot$2 & 0$\cdot$9 & 0$\cdot$5 & 0$\cdot$1 & 0$\cdot$8 & 0$\cdot$3 & 0$\cdot$8 & 0$\cdot$5 & 0$\cdot$1 & 0$\cdot$4 & 0$\cdot$3 & 0$\cdot$07 \\
maxbias & 0.814 & 1.171 & 1.566 & 1.016 & 1.256 & 1.573 & 1.482 & 1.612 & 1.610 & 2.122 & 2.084 & 1.715 \\ 
\end{tabular}
\end{adjustbox}
\end{flushleft}
\label{tab:1}
\end{table}

\begin{table}[htbp!]
\caption{Comparative performance in covariance matrix estimation in the simulation study. The performance across 20 simulation replicates is reported in terms of running time per replicate, operator norm error , mean squared error ($\times 10^2$), average absolute bias ($\times 10^2$), and maximum absolute bias for different combinations of $p$, $k$, and $g$ with $n = 200$.}
\begin{flushleft} 
\begin{adjustbox}{width = 1\textwidth}
\small
\begin{tabular}{ccccccccccccc} \toprule
{\bf p} & \multicolumn{3}{c}{{\bf{252}}} & \multicolumn{3}{c}{{\bf{504}}} & \multicolumn{3}{c}{{\bf{1008}}} & \multicolumn{3}{c}{{\bf{2016}}} \\ 
{\bf k} & \multicolumn{3}{c}{{\bf{6}}} & \multicolumn{3}{c}{{\bf{12}}} & \multicolumn{3}{c}{{\bf{24}}} & \multicolumn{3}{c}{{\bf{36}}} \\
\cmidrule(lr){2-4} \cmidrule(lr){5-7} \cmidrule(lr){8-10} \cmidrule(lr){11-13} \\
{\bf g} & {\bf 1} & {\bf 3} & {\bf 6} & {\bf 1} & {\bf 3} & {\bf 6} & {\bf 1} & {\bf 3} & {\bf 6} & {\bf 1} & {\bf 3} & {\bf 6}  \\
\hline
error & 18$\cdot$17 (0$\cdot$39) & 19$\cdot$63 (0$\cdot$04) & 20$\cdot$64 (0$\cdot$00) & 21$\cdot$90 (0$\cdot$34) & 22$\cdot$50 (0$\cdot$26) & 24$\cdot$08 (0$\cdot$00) & 27$\cdot$52 (0$\cdot$47) & 29$\cdot$41 (0$\cdot$016) & 30$\cdot$42 (0$\cdot$00) & 36$\cdot$77 (1$\cdot$57) & 39$\cdot$84 (0$\cdot$71) & 41$\cdot$44 (0$\cdot$00) \\
time & 106$\cdot$12 & 35$\cdot$96 & 18$\cdot$21 & 115$\cdot$22 & 39$\cdot$31 & 19$\cdot$91 & 153$\cdot$54 & 77$\cdot$01 & 38$\cdot$90 & 378$\cdot$80 & 155$\cdot$26 & 78$\cdot$48  \\
mse & 0$\cdot$04 & 0$\cdot$09 & 0$\cdot$49 & 0$\cdot$03 & 0$\cdot$08 & 0$\cdot$27 & 0$\cdot$02 & 0$\cdot$04 & 0$\cdot$13 & 0$\cdot$02 & 0$\cdot$03 & 0$\cdot$06 \\
avgbias & 0$\cdot$8 & 0$\cdot$6 & 0$\cdot$5 & 0$\cdot$6 & 0$\cdot$5 & 0$\cdot$3 & 0$\cdot$5 & 0$\cdot$4 & 0$\cdot$1 & 0$\cdot$4 & 0$\cdot$3 & 0$\cdot$7 \\
maxbias & 0.596 & 1.017 & 1.387 & 0.722 & 1.081 & 1.375 & 1.081 & 1.123 & 1.424 & 1.716 & 1.582 & 1.479 \\
\end{tabular}
\end{adjustbox}
\end{flushleft}
\label{tab:2}
\end{table}

\begin{figure}[htbp!]
\centering
\begin{subfigure}{.4\textwidth}
\centering
\includegraphics[width = 0.9\linewidth]{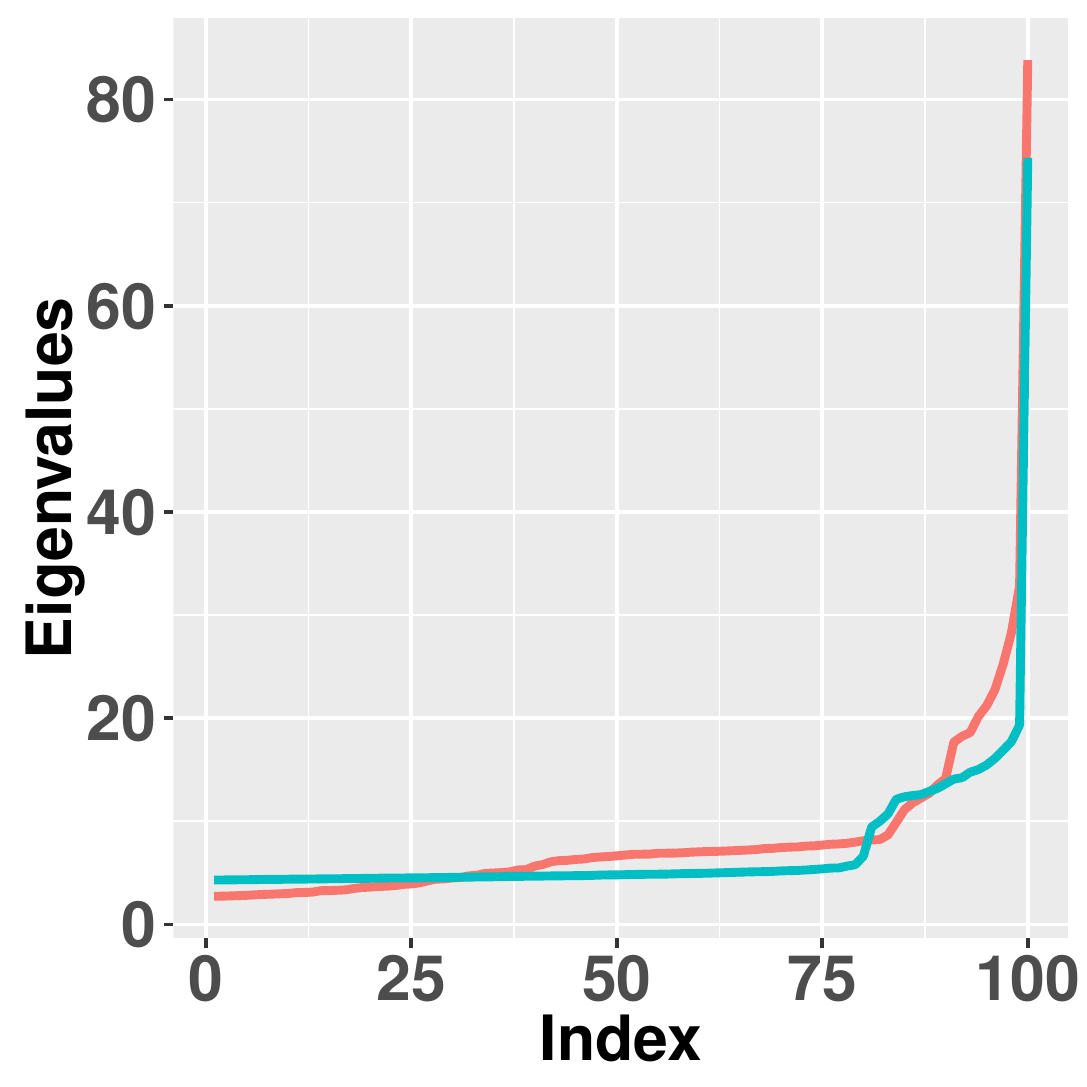}
\caption{$p = 10,000$} 
\label{fig:sub5}
\end{subfigure} 
\begin{subfigure}{.4\textwidth} 
\centering
\includegraphics[width = 0.9\linewidth]{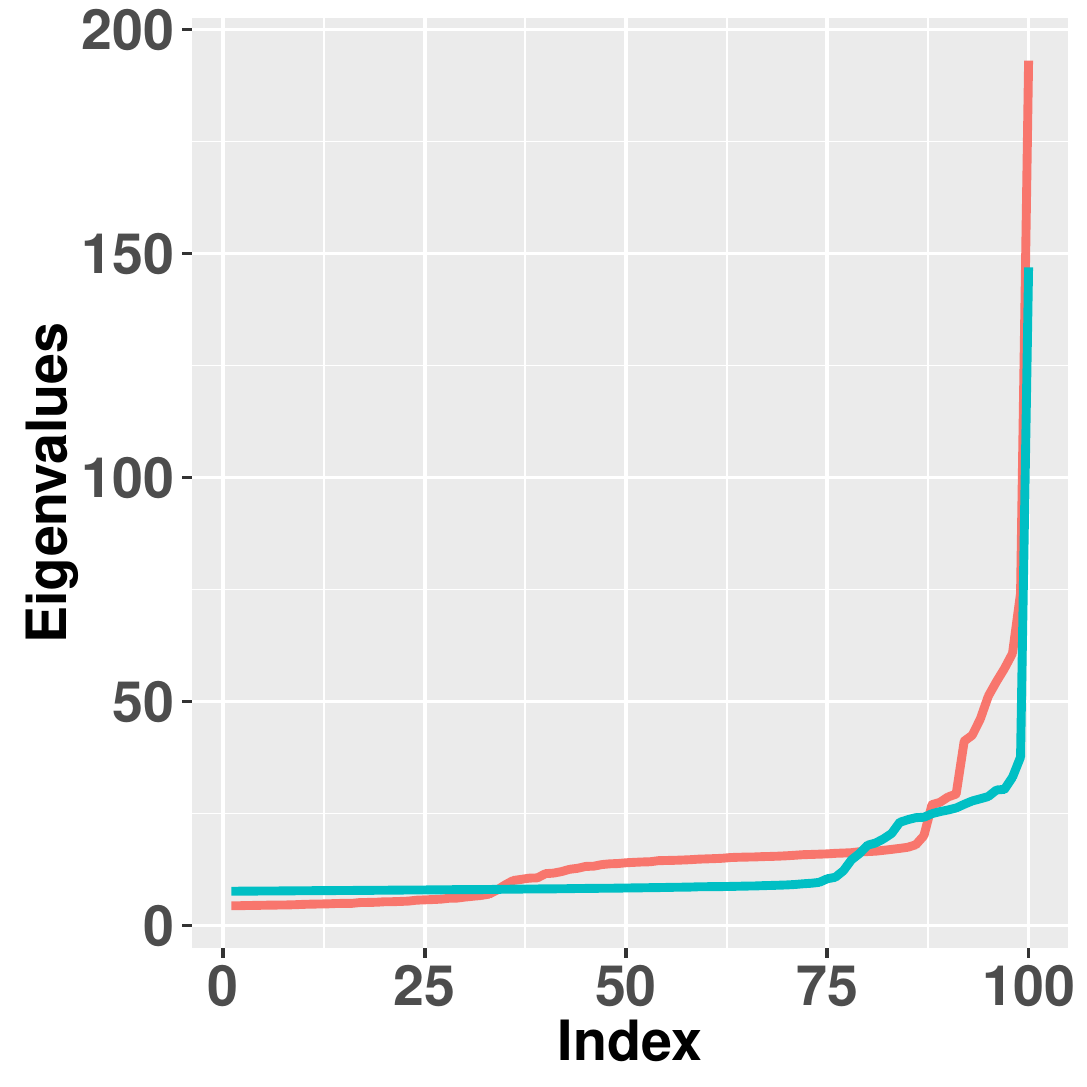} 
\caption{$p = 20,000$} 
\label{fig:sub6}
\end{subfigure} 
\caption{Eigenvalue comparisons of the covariance matrix estimators obtained for $g = 10$ (red) and $g = 20$ (blue) in the simulation study. The x-axis indexes the 100 leading eigenvalues obtained by eigendecomposition of the estimated covariance matrices. The y-axis denotes the magnitudes of eigenvalues associated with the respective index. } 
\label{fig:ev} 
\end{figure}

In our final simulation, we estimate covariance matrices for $p \asymp 10^4$. We report average, best, and worst performances of two estimators for $g \in \{10,20\}$ across $20$ simulation replicates in terms of operator norm errors in Table \ref{tab:3}.
We found that it is impossible to obtain an estimate of the covariance matrix without implementing our method. The entries ``Fail" correspond to the case $g = 1$ where estimates $\Sigma$ on a single core cannot be obtained due to substantial demands on time and memory. By implementing our algorithm for $g = 10,20$, we were able to obtain estimators that do reasonably well in terms of operator norm error and achieve substantial computational gains with increasing number of groups. These results show that there is no hope of estimating the original covariance matrix in applications where $p$ is massive. The approach would provide one with a working estimator of the covariance matrix in such applications. 
\begin{table}[htbp]
\caption{Comparative performance for covariance matrix estimation in a simulation study where $p \asymp 10^4$.  Running times per replicate, average (avgError), best (minError), and worst (maxError) performance reported in terms of operator norm errors with standard errors in parentheses}
\begin{flushleft}
\begin{adjustbox}{width = 1\textwidth}
\tiny
\begin{tabular}{ccccccc} \toprule
{\bf p} & \multicolumn{3}{c}{{\bf{10000}}} & \multicolumn{3}{c}{{\bf{20000}}} \\
{\bf k} & \multicolumn{3}{c}{{\bf{100}}} & \multicolumn{3}{c}{{\bf{200}}} \\
\cmidrule(lr){2-4} \cmidrule(lr){5-7}
{\bf g} & {\bf 1} & {\bf 10} & {\bf 20} & {\bf 1} & {\bf 10} & {\bf 20} \\ 
\hline
avgError & Fail & 46$\cdot$81 (0$\cdot$11) & 47$\cdot$28 (0$\cdot$09) & Fail & 49$\cdot$35 (0$\cdot$16) & 51$\cdot$39 (0$\cdot$11) \\ 
maxError & Fail & 47$\cdot$30 & 47$\cdot$37 & Fail & 49$\cdot$65 & 52$\cdot$39 \\ 
minError & Fail & 46$\cdot$62 & 47$\cdot$06 & Fail &  49$\cdot$31 & 50$\cdot$11 \\ 
Time & Fail & 1626 & 998 & Fail & 2234 & 1276 \\
\end{tabular}
\end{adjustbox}
\end{flushleft}
\label{tab:3}
\end{table} 
In Figure \ref{fig:ev}, we plot the $100$ leading eigenvalues of the two estimators for comparison. We see that the estimated leading eigenvalues obtained via eigendecomposition of $\hat{\Sigma}^{(10)}$ and ${\hat{\Sigma}^{(20)}}$ are comparable. 
\section{Incidence of Statin-induced Myotoxicity Application} \label{realdata} 
Statins are a class of lipid-lowering medications that control the production of cholesterol in the human body. High cholesterol levels are associated with cardiovascular disease risk, which is one of the leading causes of death globally. Statins are widely prescribed and have been shown to have beneficial effects in a broad range of patients in the reduction of cholesterol levels. However, statins are associated with several adverse side effects such as muscle problems, an increased risk of diabetes, and increased liver enzymes in the blood due to liver damage.  \cite{mangravite2013statin} studied the effects of {\textit{in vitro}} statin exposure on gene expression levels, in lymphoblastoid cell lines derived from 480 participants in genomic study. For each participant, $5,509$ of $10,195$ expressed genes had a significant interaction with simvastatin exposure. The magnitude of change in expression across significant genes is small with $1,952$ genes exhibiting greater than or equal to $10\%$ change in expression, and only 21 genes exhibiting greater than or equal to $50\%$ change in expression \cite{mangravite2013statin}. 

\begin{figure}[htp!]
\centering
\includegraphics[width=0.4\linewidth]{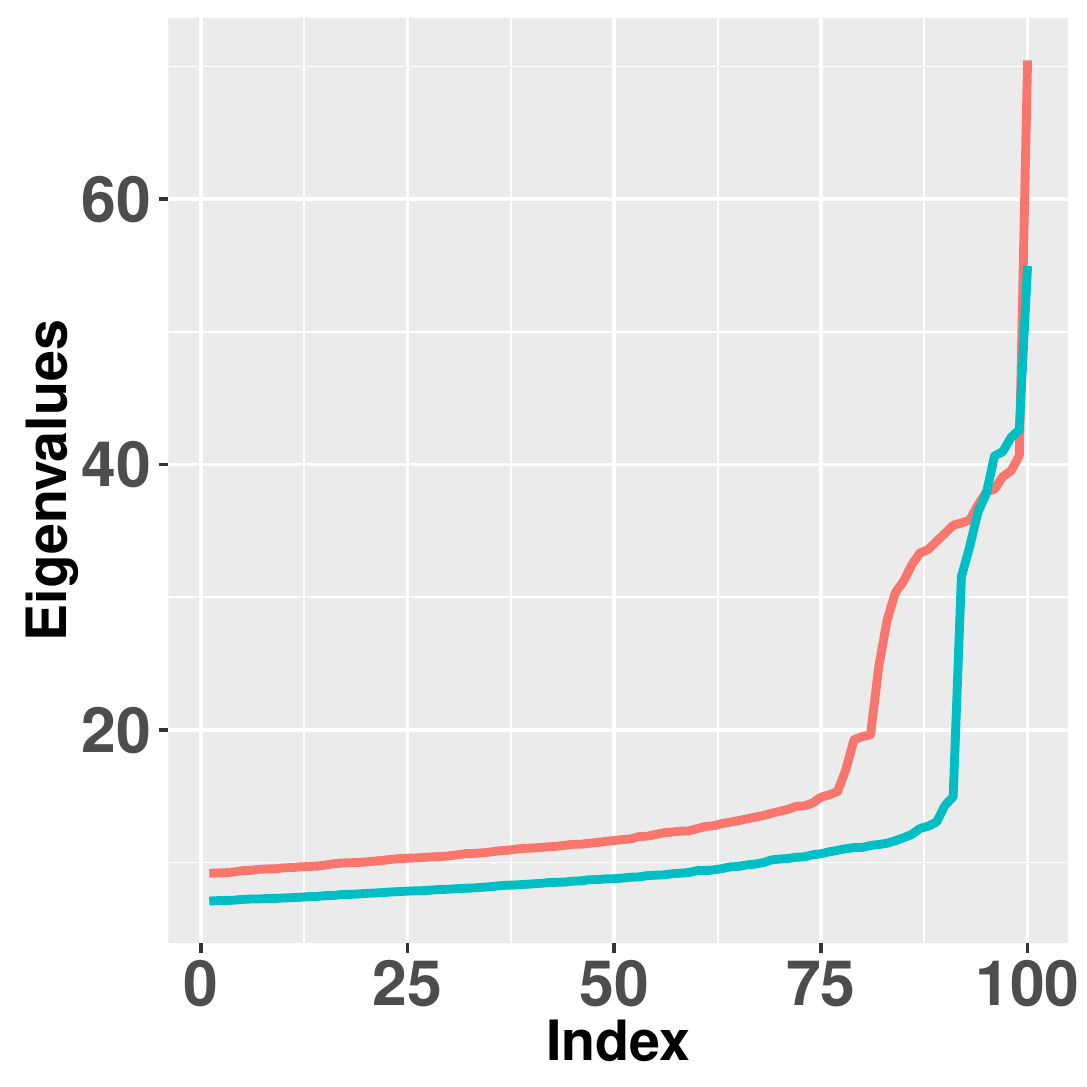}
\caption{Eigenvalue comparisons of the covariance matrix estimators for $g = 10$ (red) and $g = 20$ (blue) obtained for gene expression data. The x-axis indexes the 100 leading eigenvalues obtained by eigendecomposition of the estimated covariance matrices. The y-axis denotes the magnitudes of eigenvalues associated with the respective index.} 
\label{fig:real} 
\end{figure}
Our interest lies in simultaneously identifying the six eQTLs by modeling the second order structure among the genes via a latent factor model. Let $Y_i$ denote $10,195$ dimensional gene expression vector for participant $i$. The values in each cell of the vector vary from -3 to 3. 
Let ${{Y}}$ denote the $480 \times 10,195$ data matrix where $p = 10,195$ and $n = 480$.  We estimate the covariance matrix to obtain ${\hat{\Sigma}}_{10}$ for ten groups and ${\hat{\Sigma}}_{20}$ for twenty groups. The posterior analyses proceed exactly as in \cite{mangravite2013statin}, but an additional step is needed to compute the adjacency matrices corresponding to the estimated covariance matrices. We ran the Gibbs sampler for 10,000 iterations with 5,000 burn-in and collected every first sample after burn-in to thin the chain. The number of factors was set to 100. The posterior mean of $\rho$ is 0.3 showing reasonable correlation among the sub-groups.  Figure \ref{fig:real} shows that the $100$ leading eigenvalues of the two estimators are comparable. To gain more insight into the estimated covariance, we threshold the entries of the correlation matrix to create an adjacency matrix of the gene-regulatory network containing 0s and 1s. Then, using the \texttt{igraph} package in \texttt{R}, we clustered the genes in this correlation network, and we found a dominant cluster of around 2,000 genes and a few smaller clusters.  A Gene Ontology enrichment analysis \cite{eden2009gorilla} on these clusters demonstrated that clusters 1 and 4  sorted by decreasing order of their sizes 
both included subsets of genes in the cluster with shared biological processes at FDR $\leq 0.05$.  This enrichment of a specific shared biological process suggests that the clusters that we have identified are biologically coherent, but more research is needed to understand what is jointly regulating the co-expressed genes.
\appendix
\section{}
\small
\subsection{Proof of Lemma \ref{rank}}\label{pfofL3}
Let $\Lambda,\Lambda^{*} \in {\mathbb{R}}^{p \times k}$ where $\Lambda^{*} = DC$ such that $D=\mbox{diag}(\Lambda^{(1)}, \ldots, \Lambda^{(g)})$ and $C^2 =  C^* \otimes I_{k_g}$, where $C^*$ is a $g \times g$ positive definite matrix such that $C^*_{mm'} = 1$ if $m=m'$ and $C^*_{mm'} = \rho$ if $m \neq m'$ for $0< \rho < 1$.  Letting $\mbox{rank}(\Lambda^{(m)}) = k_g, \; m = 1, \ldots, g$, it is enough to show that $A = \Lambda\Lambda^{\T}$ and $A^{*} = \Lambda^{*}\Lambda^{*\T}$ have the same rank.
Observe that $\mbox{rank}(A) = \mbox{rank}(\Lambda) = k$ and  $C$ is invertible since \begin{align*}
\mbox{det}(C)  & = \mbox{det}(C^{*} \otimes I_{k/2}) \tag{2} 
= \mbox{det}(C^{*})^{k/2}\;\mbox{det}(I_{k/2})^{k} 
= (1 + (k-1)\rho)(1 - \rho)^{k-1} > 0.  
\end{align*}
We will show $\mbox{rank}(A^{*}) = \mbox{rank}(\Lambda^{*}) = k$. This follows from the fact that $\mbox{rank}(\Lambda^{*})  = \mbox{rank}(DC) = \mbox{rank}(D) = \sum_{j=1}^g \mbox{rank}(\Lambda^{(g)}) 
 = \mbox{k}$. 

\subsection{Proof of Lemma \ref{trace}}

Let $\Lambda_0 = [\Lambda_{01}, \ldots, \Lambda_{0k}]$ be such that $\Lambda_{0h} \in \ell_0[s;p]$ for $h = 1, \ldots, k$ and $\Lambda^{*}$ be the same as defined in the proof of Lemma \ref{rank} in Appendix \ref{pfofL3}. Then
\begin{align*} 
\mathrm{pr}(\lvert\mbox{tr}(\Lambda^{*}\Lambda^{*\T}) - \mbox{tr}(\Lambda_0\Lambda_0^{\T})\rvert < \epsilon) & = \mathrm{pr}\bigg\{\left|{\displaystyle\sum\limits_{m=1}^{g}}{\displaystyle\sum\limits_{h=1}^{k_g}}{\lVert {\Lambda_h^{(m)}}\rVert}^{2} - {\displaystyle\sum\limits_{h=1}^{k}{\lVert{\Lambda_{0h}}\rVert}^{2}}\right| < \epsilon\bigg\} .\tag{A.1}
\end{align*}
To lower bound (A.1), we first obtain a lower bound conditioned on the hyper parameter $\tau = (\tau_1. \ldots, \tau_k)$ and then intergrate out $\tau$: 
$\mathrm{pr}\bigg\{\left|{\displaystyle\sum\limits_{m=1}^{g}}{\displaystyle\sum\limits_{h=1}^{k_g}}{\lVert {\Lambda_h^{(m)}}\rVert}^{2} - {\displaystyle\sum\limits_{h=1}^{k}{\lVert{\Lambda_{0h}}\rVert}^{2}}\right| < \epsilon\bigg\}$
\small\begin{align*} 
& \ge \bigintss_{({\mathbb{R}}^{+})^{k}}\mathrm{pr}\bigg\{\left|{\displaystyle\sum\limits_{m=1}^{g}}{\displaystyle\sum\limits_{h=1}^{k_g}}{\lVert {\Lambda_{hS_0^c}^{(m)}}\rVert}^{2}\right| < \epsilon/2 \mid\tau\bigg\}pr\bigg\{\left|{\displaystyle\sum\limits_{m=1}^{g}}{\displaystyle\sum\limits_{h=1}^{k_g}}{\lVert {\Lambda_{hS_{0h}}^{(m)}}\rVert}^{2} - {\displaystyle\sum\limits_{h=1}^{k}{\lVert{\Lambda_{0hS_{0h}}}\rVert}^{2}}\right| < \epsilon/2 \mid \tau\bigg\} {\displaystyle\prod\limits_{h=1}^{k}g_h(\tau_h)}d\tau_h\\
& \ge \bigintss_{({\mathbb{R}}^{+})^{k}}\mathrm{pr}\big\{\lvert\Lambda_{jh}^{(m)}/\sqrt{\tau_h}\rvert < \delta/\sqrt{\tau_h}\; \mbox{for all}\; j \in S_{0h}^c \mid \tau_h\big\}\;pr\bigg\{\left|{\displaystyle\sum\limits_{m=1}^{g}}{\displaystyle\sum\limits_{h=1}^{k_g}}{\lVert {\Lambda_{hS_{0h}}^{(m)}}\rVert}^{2} - {\displaystyle\sum\limits_{h=1}^{k}{\lVert{\Lambda_{0hS_{0h}}}\rVert}^{2}}\right| < \epsilon/2 \mid \tau\bigg\}{\displaystyle\prod\limits_{h=1}^{k}g_h(\tau_h)}d\tau_h\\
& \approx \bigintss_{({\mathbb{R}}^{+})^{k}}\bigg\{\displaystyle\prod\limits_{j \in S_0^c}{\exp{}({\log{}({\delta/\sqrt{\tau_h}})})}\bigg\}\;\mathrm{pr}\bigg\{\left|{\displaystyle\sum\limits_{m=1}^{g}}{\displaystyle\sum\limits_{h=1}^{k_g}}{\lVert {\Lambda_{hS_{0h}}^{(m)}}\rVert}^{2} - {\displaystyle\sum\limits_{h=1}^{k}{\lVert{\Lambda_{0hS_{0h}}}\rVert}^{2}}\right| < \epsilon/2 \mid \tau\bigg\}{\displaystyle\prod\limits_{h=1}^{k}g_h(\tau_h)}d\tau_h \tag{A.2}
\end{align*}
where $S_{0h} = \mbox{supp}(\Lambda_{0h})$, $\delta = \sqrt{\epsilon}/\sqrt{2k(p-s)}$ and (A.2) follows by noting that $\Lambda_{jh}/\sqrt{\tau_h} \mid \tau_{h} \sim t_{\nu}$.
Fix $a < 1$ , $0 < c < d$ and define $\mathcal{B} \subset ({\mathbb{R}}^{+})^{k}$ such that 
\begin{align*}
\mathcal{B} = \bigg\{ \tau : \frac{\delta}{da^{s/{p-s}}} < \sqrt{\tau_h} < \frac{\delta}{ca^{s/{p-s}}} \quad \mbox{for every}\,  h = 1, \ldots, k\bigg\} \tag{A.3}
\end{align*}
Thus,
\small\begin{multline*}
\mathrm{pr}\bigg\{\left|{\displaystyle\sum\limits_{m=1}^{g}}{\displaystyle\sum\limits_{h=1}^{k_g}}{\lVert {\Lambda_h^{(m)}}\rVert}^{2} - {\displaystyle\sum\limits_{h=1}^{k}{\lVert{\Lambda_{0h}}\rVert}^{2}}\right| < \epsilon\bigg\} \ge \underset{\tau \in \mathcal{B}}{\inf}\;\mathrm{pr}\bigg\{\left|{\displaystyle\sum\limits_{m=1}^{g}}{\displaystyle\sum\limits_{h=1}^{k_g}}{\lVert {\Lambda_{hS_{0h}}^{(m)}}\rVert}^{2} - {\displaystyle\sum\limits_{h=1}^{k}{\lVert{\Lambda_{0hS_{0h}}}\rVert}^{2}}\right| < \epsilon/2 \mid \tau\bigg\} \\ \bigintss_{\mathcal{B}}\bigg\{\displaystyle\prod\limits_{j \in S_0^c}{\exp{}({\log{}({\delta/\sqrt{\tau_h}})})}\bigg\}{\displaystyle\prod\limits_{h=1}^{k}g_h(\tau_h)}d\tau_h. \tag{A.4}
\end{multline*}
It is possible to obtain a tight bound on the first probability term on right hand side of (A.4) since $\Lambda_{hS_{0h}}^{(m)}$ has a high concentration around the truth $\Lambda_{0hS_{0h}}$ for all $h = 1, \ldots, k, \; m = 1,\ldots,g$ for $ \mid{S_{0h}}\mid = s \ll p$.
For $\tau \in \mathcal{B}$, the term inside the integral in (A.4) can be bounded below as follows: 
\begin{align*} 
\bigg\{\displaystyle\prod\limits_{j \in S_0^c}{\exp{}({\log{}({\delta/\sqrt{\tau_h}})})}\bigg\} \ge \exp{}(s\log{}a) \tag{A.5}
\end{align*}
To obtain a lower bound for pr$(\mathcal{B})$, we define
\begin{align*}
\mathcal{B^{*}} = \bigg\{\delta \in {({\mathbb{R}}^{+})^{k}} : \frac{\delta}{da^{s/{p-s}}} < \sqrt{\delta_1} < \frac{\delta}{ca^{s/{p-s}}}, \sqrt{\delta_h} \le 1\; \mbox{for every}\; h \ge 2 \bigg\}
\end{align*}
Since $\delta_{\ell}\; \ell = 1, \ldots, k$ are independent, we have 

\begin{equation*}
\mathrm{pr}({\mathcal{B}}) \ge \mathrm{pr}({\mathcal{B}}^{*}) = C(a_1, a_2, k) \bigg\{\frac{\delta}{a^{s/(p-s)}}\bigg\}^{2}(1/c^2 - 1/d^2) \tag{A.6} 
\end{equation*} 
where $C(a_1, a_2,k) = 1/\Gamma(a_1)\{1/\Gamma(a_2)\}^{k-1}(1/a_1a_2^{k-1})$. 
Finally, (A.5) and (A.6) substituted into (A.4) give us 
\begin{multline*}
\mathrm{pr}(\lvert \mbox{tr}(\Lambda^{*}\Lambda^{*\T}) - {\mbox{tr}}(\Lambda_0\Lambda_0^{\T}))\rvert < \epsilon) \ge \underset{\tau \in \mathcal{B}}{\inf}\;\mathrm{pr}\bigg\{\left|{\displaystyle\sum\limits_{m=1}^{g}}{\displaystyle\sum\limits_{h=1}^{k_g}}{\lVert {\Lambda_{hS_{0h}}^{(m)}}\rVert}^{2} - {\displaystyle\sum\limits_{h=1}^{k}{\lVert{\Lambda_{0hS_{0h}}}\rVert}^{2}}\right| < \epsilon/2 \mid \tau\bigg\} \\ \exp{}\big[-C\{s\log{}(1/a) + \log{}(2k(p-s)/\epsilon)\}\big]
\end{multline*}
for some constant $C > 0$. 

\section{}
\subsection{Posterior Computation: Obtaining posterior sub-estimates} \label{posterior}
We propose a straightforward Gibbs sampler for posterior computation after augmenting the latent factors to incorporate a dependence structure across estimates obtained from different cores. The Gibbs sampler, using the multiplicative Gamma process prior \cite{bhattacharya2011sparse}, adapted to our framework cycles through the following steps on each of the cores.
\begin{enumerate}
\item Sample $X_i, i = 1, \dots, n$ from conditionally independent Gaussian posteriors 
\begin{multline*} 
X_i \mid {\mbox{rest}} \sim \mbox{N} \bigg[\bigg\{{\rho{\displaystyle\sum\limits_{m=1}^{g}}{\Lambda^{(m)\T}}{\Sigma^{-(m)}}{\Lambda^{(m)}}} + I_{p_g}\bigg\}^{-1}{\sqrt{\rho}}{\displaystyle\sum\limits_{m=1}^{g}}{\Lambda^{(m)\T}}{\Sigma^{-(m)}}{Q_i^{(m)}},  \\ \bigg\{{\rho{\displaystyle\sum\limits_{m=1}^{g}}{\Lambda^{(m)\T}}{\Sigma^{-(m)}}{\Lambda^{(m)}}} + I_{p_g}\bigg\}^{-1} \bigg]
\end{multline*}
where {\small $Q_i^{(m)} = Y_i^{(m)} - \sqrt{1 - \rho}\;{\Lambda^{(m)}}Z_i^{(m)}$} and {\small $\Sigma^{-(m)} = {\Sigma^{(m)}}^{-1}$}. Observe how the update for $X_i$ utilises the information stored in other posterior quantities by summing them across all $g$ machines. 
\item Sample ${{Z_i^{(m)} \mid {\text{rest}}}}, i = 1, \dots, n, \; m = 1, \dots, g$ from conditionally independent Gaussian posteriors
\begin{multline*} 
Z_i^{(m)}\mid {\mbox{rest}} \sim \mbox{N} \bigg[\bigg\{(1 - \rho){\Lambda^{(m)\T}}{\Sigma^{-(m)}}{\Lambda^{(m)}} + I_{k_g}\bigg\}^{-1}{\sqrt{1 - \rho}}\;{\Lambda^{(m)\T}}{\Sigma^{-(m)}}{R_i^{(m)}}, \\ \bigg\{(1 - \rho){\Lambda^{(m)\T}}{\Sigma^{-(m)\T}}{\Lambda^{(m)}} + I_{k_g}\bigg\}^{-1}\bigg]
\end{multline*}
where $R_i^{(m)} = Y_i^{(m)} - \sqrt{\rho}\;{\Lambda^{(m)}}X_i$. In contrast to the posterior update of $X_i$, the component which is shared across $g$ machines, the posterior update of $Z_i^{(m)}$ utilizes the posterior quantities obtained on the $m$-th machine. 
\item Update ${{\eta_i^{(m)} \mid {\text{rest}}}}, i = 1, \ldots, n$
\begin{center} 
$\eta_i^{(m)} \mid {\text{rest}} = X_i \mid {\mbox{rest}} + Z_i^{(m)} \mid {\mbox{rest}}$
\end{center} 
\item Let $\lambda_j^{(m)\T}, j = 1, \ldots, p_g,\; m = 1, \ldots,g$ denote the jth row of $\Lambda^{(m)}$, then $\lambda_j^{(m)}$s have independent conidtionally conjugate posteriors, 

\begin{multline*}
{\lambda_j^{(m)}} \mid {\mbox{rest}} \sim \mbox{N} \bigg[\bigg\{{D_{j}^{-1}} + \sigma_{j}^{-2}{\eta^{(m)\T}}{\eta^{(m)}}\bigg\}^{-1}{\eta^{(m)\T}}{\sigma_{j}^{-2}}{y_{j}^{(m)}}, \bigg\{{D_{j}^{-1}} + {\sigma_{j}^{-2}}{\eta^{(m)\T}}{\eta^{(m)}} \bigg\}^{-1}\bigg]
\end{multline*}
where {\small${D_{j}^{-1}} = \text{diag}(\phi_{j1}\tau_{1}, \ldots, {\phi_{jk_g}}\tau_{k_g})$}, $\eta^{(m)} = (\eta_1^{(m)}, \ldots, \eta_n^{(m)})^{\T}$ and $y_j^{(m)} = (y_{1j}^{(m)}, \ldots, y_{nj}^{(m)})^{\T}$.
\item Sample ${\phi_{jh}}, j = 1, \ldots, p_g, \; h = 1, \ldots, k_g$ across all machines from conditionally independent Gamma posteriors.
\begin{center} 
${\phi_{jh}} \mid {\text{rest}} \sim \Gamma\bigg\{\dfrac{\nu}{2} + 1, \;\frac{\nu + \tau_h^{(m)}{\lambda_{jh}^{(m)}}^{2}}{2}\bigg\}$
\end{center}

\item Sample $\delta_1$ from conditionally independent Gamma posteriors. 
\begin{center} 
${\delta_1} \mid {\mbox{rest}} \sim \Gamma\bigg\{\frac{p_gk_g}{2} + a_1, 1 + {\displaystyle\sum\limits_{h=1}^{k_g}{\displaystyle\prod\limits_{l = 2}^{h}{\delta_l}}}{\displaystyle\sum\limits_{j=1}^{p_g}{\phi_{jh}}{\lambda_{jh}^{(m)}}^{2}}\bigg\}$
\end{center}
\item Sample $\delta_h$ for $h\ge 2$ from conditionally independent Gamma posteriors
\begin{center}
$\delta_h \mid {\text{rest}} \sim \Gamma\bigg\{a_2 + \frac{p_g}{2}(k_g - h + 1), 1 + \frac{1}{2}{\displaystyle\sum\limits_{l = h}^{k_g}{\tau_l^{(h)}}}{\displaystyle\sum\limits_{j = 1}^{p_g}{\phi_{jl}}{\lambda_{jl}^{(m)}}^{2}}\bigg\}$
\end{center}
where $\tau_l^{(h)} = \displaystyle\prod\limits_{t = 1, t \ne h}^{l}{\delta_t}$ for $h  = 1, \dots, K$. 
\normalsize
\end{enumerate}

\bibliographystyle{plain}
\bibliography{references}

\end{document}